\documentclass[a4paper]{jpconf}
\usepackage{graphicx}
\begin{document}
\title{Stochastic approach to the numerical solution of the non-stationary Parker's transport equation}

\author{A. Wawrzynczak$^{1}$, R. Modzelewska$^{2}$, A. Gil$^{2}$}

\address{$^{1}$Institute of Computer Sciences, Siedlce University, Poland, $^{2}$Institute of Mathematics and Physics, Siedlce University, Poland.}

\ead{awawrzynczak@uph.edu.pl, renatam@uph.edu.pl, gila@uph.edu.pl,}

\begin{abstract}
We present the newly developed stochastic model of the galactic cosmic ray (GCR) particles transport in the heliosphere. Mathematically Parker transport equation (PTE) describing non-stationary transport of charged particles in the turbulent medium is the Fokker-Planck type. It is the second order parabolic time-dependent 4-dimensional (3 spatial coordinates and particles energy/rigidity) partial differential equation. It is worth to mention that, if we assume the stationary case ($\partial f/\partial t =0$) it remains as the 3-D parabolic type problem with respect to the particles rigidity $R$. If we fix the energy ($\partial f/\partial R=0$) it still remains as the 3-D parabolic type problem with respect to time. The proposed method of numerical solution is based on the solution of the system of stochastic differential equations (SDEs) being equivalent to the Parker's transport equation. We present the method of deriving from PTE the equivalent SDEs  in the heliocentric spherical coordinate system for the backward approach. The obtained stochastic model of the Forbush decrease of the GCR intensity is in an agreement with the experimental data. The advantages and disadvantages of the forward and the backward solution of the PTE are discussed.
\end{abstract}

\section{Introduction}
Many problems in physics, finance or biology can be represented as models of the diffusive transport processes described by the Fokker-Planck type equations (FPE). The difficulty of the numerical solution of this type equation increases with the problem dimension. Reason is the instability of the numerical schemes like finite-differences and finite-volume in the higher dimensions. To ensure the scheme stability and convergence the density of numerical grid must be improved, increasing the computational complexity. To overcome this problem the stochastic methods can be applied (e.g.\cite{Zhang1999},\cite{Gerv}). In this approach the individual particle motion is described as a Markov stochastic process, and the system evolves probabilistically. Accordingly, the FPE can be solved by the corresponding stochastic differential equations (SDEs) (e.g.~\cite{Ga}).\\
We apply stochastic methodology to model the galactic cosmic rays (GCR) transport in the heliosphere.  During the propagation through the heliosphere GCR particles are modulated by the solar wind and heliospheric magnetic field (HMF). Modulation of the GCR is a result of action of four main processes: convection by the solar wind, diffusion on irregularities of HMF, particles drifts in the non-uniform magnetic field and adiabatic cooling (e.g. \cite{Mo}). Transport of the GCR particles in heliosphere is described by the Parker transport equation (PTE) \cite{Pa} being the second order parabolic type partial differential equation:
\begin{eqnarray} \label{ParkerEq}
\frac{\partial f}{\partial t}=\vec{\nabla}\cdot (K_{ij} ^{S}\cdot \vec{\nabla}f)-(\vec{v}_{d}+\vec{U})\cdot \vec{\nabla} f+\frac{R}{3}(\vec{\nabla} \cdot \vec{U})\frac{\partial f}{\partial R},
\end{eqnarray}
where $f=f(\vec{r}, R, t) $ is an omnidirectional distribution function of three spatial coordinates $\vec{r}=r(r,\theta,\varphi)$, particles rigidity $R$ and time $t$;  $\vec{U}$  is solar wind velocity, $\vec{v}_{d}$  the drift velocity, and $K_{ij} ^{S}$ is the symmetric part of the diffusion tensor of the GCR particles.\\
Based on the stochastic approach we model the short time variation of the GCR intensity, called the Forbush decrease (Fd) \cite{Forbush54}.
The Fd occurs as an occasional decrease in GCR intensity recorded on the Earth surface by the neuron monitors.The Fds follow the occurrence of the solar flares and intensive solar coronal mass ejecta (CME) \cite{Cane00}. There can be distinguished two types of the Fds: 1) sporadic- being the result of the shock waves and the magnetic clouds appearing in the interplanetary space, as the result of the solar flares on the Sun; 2) recurrent type Fd connected with the corotating interaction regions (CIR) appearing in the interplanetary space in connection with the solar rotation.
\section{Stochastic approach}
To model the GCR transport by the stochastic methods the corresponding SDEs must be derived. Firstly, the PTE (Eq.\ref{ParkerEq}) must be transformed to the standard form of the FPE which, depending on integration's direction, can be generally expressed in two forms \cite{Ga}:\\
time-forward:
\begin{eqnarray}
\label{forwardFPE}
\frac{\partial F}{\partial t}=\sum _{i}\frac{\partial}{\partial x_{i}}(A_{i}\cdot F)+\frac{1}{2}\sum _{i,j}\frac{\partial^{2}}{\partial x_{i}\partial x_{j}}(B_{ij}B_{ij}^{T}\cdot F),
\end{eqnarray}
time-backward:
\begin{eqnarray}
\label{backwardFPE}
\frac{\partial F}{\partial t}=\sum _{i} A_{i} \frac{\partial F}{\partial x_{i}}+\frac{1}{2}\sum _{i,j} B_{ij}B_{ij}^{T} \frac{\partial^{2}F}{\partial x_{i}\partial x_{j}}.
\end{eqnarray}
Corresponding to Eqs.~\ref{forwardFPE} and \ref{backwardFPE}  SDE can be written as (e.g. \cite{Ga}):
\begin{eqnarray}
d\vec{r}=\vec{A_{i}}\cdot dt+B_{ij}\cdot d\vec{W},
\end{eqnarray}
where $\vec{r}$ is the trajectory of individual pseudoparticle in the phase space and $dW_{i}$ is the Wiener process, usually written as $dW_{i}=\sqrt{dt}\cdot dw_{i}$, $dw_{i}$ is the randomly fluctuating term having Gaussian distribution. \\
In both, forward and backward cases, first we start from some initial position in space and time and integrate along the pseudoparticles trajectories until they reach the boundary. The choice between the forward and backward integration depends on the problem that has to be solved. In the case of the GCR propagation in the forward approach particles should be initialized at various points at the boundary, where the GCR particles enter the heliosphere. Then, its trajectory should be traced until they arrive the point of interest, e.g. Earth orbit. To obtain the reasonable statistic a huge number of particles should be initialized because most of them do not reach the Earth orbit.  The backward approach is more efficient for the GCR propagation in the heliosphere, because it reduces the number of 'useless' particles.  In the backward integration particles are initialized at the Earth orbit and traced backward in time until they reach the heliosphere's boundary (in this paper assumed at 100 AU, Fig.~\ref{fig:Model}). The particle distribution function can be obtained by averaging over the entrance points, $f(\vec{r}, R)=\frac{1}{N}\sum_{n=1}^{N}f_{LIS}(R)$,
where $f_{LIS}(R)$ is the cosmic ray spectrum at the outer boundary taken as in \cite{Webber} and $R$ is the rigidity of the
$n^{th}$ particle at the entrance point.\\
The PTE (Eq.\ref{ParkerEq}) in the 3-D spherical coordinate system $(r,\theta,\varphi)$  can be written as time-backward FPE diffusion equation:
\begin{eqnarray}
\label{backwardParker}
\frac{\partial f}{\partial t}&=&A_{1}\frac{\partial^{2} f}{\partial r^{2}}+A_{2}\frac{\partial^{2} f}{\partial \theta^{2}}+A_{3}\frac{\partial^{2} f}{\partial \varphi^{2}}+A_{4}\frac{\partial^{2} f}{\partial r \partial \theta}+A_{5}\frac{\partial^{2} f}{\partial r \partial \varphi}+A_{6}\frac{\partial^{2} f}{\partial \theta \partial \varphi}+A_{7}\frac{\partial f}{\partial r}+\nonumber\\&+&A_{8}\frac{\partial f}{\partial \theta}+A_{9}\frac{\partial f}{\partial \varphi}+A_{10}\frac{\partial f}{\partial R}
\end{eqnarray}
with following coefficients:\\
$A_{1}=K_{rr}^{S}, A_{2}=\frac{K_{\theta \theta}^{S}}{r^{2}}, A_{3}=\frac{K_{\varphi \varphi}^{S}}{r^{2}sin^{2} \theta}, A_{4}=\frac{2K_{r \theta }^{S}}{r}, A_{5}=\frac{2K_{r \varphi}^{S}}{r sin \theta}, A_{6}=\frac{2K_{\theta \varphi}^{S}}{r^{2}sin \theta}$\\
$A_{7}=\frac{2}{r}K_{rr}^{S}+\frac{\partial K_{rr}^{S}}{\partial r}+\frac{ctg\theta}{r}K_{\theta r}^{S}+\frac{1}{r}\frac{\partial K_{\theta r}^{S}}{\partial \theta}+\frac{1}{r sin \theta}\frac{\partial K_{\varphi r}^{S}}{\partial \varphi}-U-v_{d,r}$\\
$A_{8}=\frac{K_{r \theta}^{S}}{r^{2}}+\frac{1}{r}\frac{\partial K_{r \theta}^{S}}{\partial r}+\frac{1}{r^{2}}\frac{\partial K_{\theta \theta}^{S}}{\partial \theta}+\frac{ctg \theta}{r^{2}}K_{\theta \theta}^{S}+\frac{1}{r^{2} sin \theta}\frac{\partial K_{\varphi \theta}^{S}}{\partial \varphi}-\frac{1}{r}v_{d,\theta}$\\
$A_{9}=\frac{K_{r \varphi}^{S}}{r^{2} sin\theta}+\frac{1}{r sin \theta}\frac{\partial K_{r \varphi}^{S}}{\partial r}+\frac{1}{r^{2}sin \theta}\frac{\partial K_{\theta \varphi}^{S}}{\partial \theta}+\frac{1}{r^{2} sin^{2} \theta}\frac{\partial K_{\varphi \varphi}^{S}}{\partial \varphi}-\frac{1}{r sin \theta}v_{d,\varphi}$\\
$A_{10}=\frac{R}{3}\nabla \cdot U$.\\
We apply the full 3D anisotropic diffusion tensor of GCR  particles $K_{ij}=K_{ij} ^{(S)}+K_{ij} ^{(A)}$ consisting of the symmetric $K_{ij} ^{(S)}$  and antisymmetric $K_{ij} ^{(A)}$  parts presented in \cite{Alania02}.  The drift velocity of GCR particles  is implemented as: $ v_{d,i}=\frac{\partial K_{ij} ^{(A)}}{\partial x_{j}}$ \cite{jokipii77}. The corresponding to Eq.~\ref{backwardParker} set of SDEs with matrix $B_{ij}$, $(i,j=r,\theta,\varphi)$ has a form (the same form can be found in \cite{Kopp2012}):

\begin{minipage}{14pc}
\begin{eqnarray}\label{SDE}
dr &=& A_{7} \cdot dt+[B \cdot dW]_{r} \nonumber \\
d \theta &=& A_{8}\cdot dt+[B \cdot dW]_{\theta}\\
d\varphi &=& A_{9}\cdot dt+[B\cdot dW]_{\varphi} \nonumber\\
dR &=& A_{10}\cdot dt.\nonumber
\end{eqnarray}
\end{minipage}\hspace{2pc}%
\begin{minipage}{18pc}
\[ B_{i,j} = \left[ \begin{array}{ccc}
 \sqrt{2A_{1}} & 0 & 0 \\
\frac{A_{4}}{\sqrt{2A_{1}}} &  \sqrt{2A_{2}-\frac{A_{4}^{2}}{2A_{1}}} & 0 \\
\frac{A_{5}}{\sqrt{2A_{1}}} & \frac{A_{6}-\frac{A_{4}A_{5}}{2A_{1}}}{B_{\theta \theta}} & \sqrt{2A_{3}-B_{\varphi r}^{2}-B_{\varphi \theta}^{2}} \end{array} \right].\]
\end{minipage}\\
The Eqs. \ref{SDE} are integrated backward in time by the Euler$-$Maruyama scheme.
As an initial condition, an empty heliosphere is assumed, as discussed in \cite{Pei2010}. Solving Eqs. \ref{SDE} in spherical coordinates the following boundary conditions are assumed: $\varphi_{i}<0 \rightarrow \varphi_{i}=\varphi_{i}+2 \pi $, $\varphi_{i}>2 \pi \rightarrow \varphi_{i}=\varphi_{i}-2\pi $, $\theta_{i}<0 \rightarrow \theta_{i}=\theta_{i}+2\pi $ and  $\theta_{i}> \pi \rightarrow \theta_{i}=\pi - |\theta_{i}| $. For the inner radial boundary we assume the reflecting boundary, $\frac{\partial f}{\partial r}=0$ at $r=0.001$ AU, with time from $0$ to $100$ days. The outher boundary is specified at radial distance $100$AU as $f(100,R,t)=f_{LIS}(R)$.

\section{Model of the short time variation of the GCR intensity}

\begin{figure}[tbp]
\begin{minipage}{18pc}
\includegraphics[width=1\hsize]{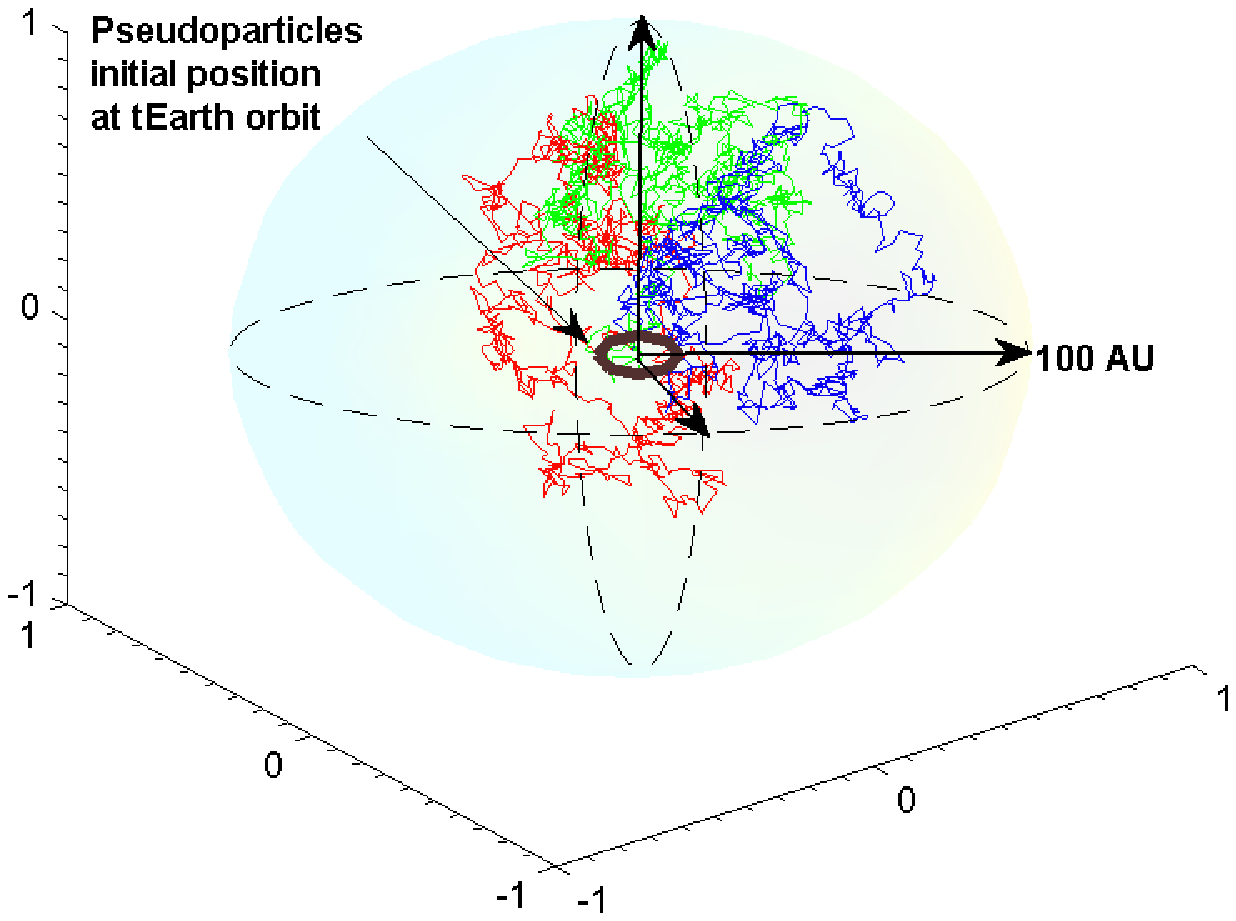}
\caption{\label{fig:Model} The sample pseudoparticles trajectories within the heliosphere.}
\end{minipage}\hspace{2pc}%
\begin{minipage}{18pc}
\includegraphics[width=1\hsize]{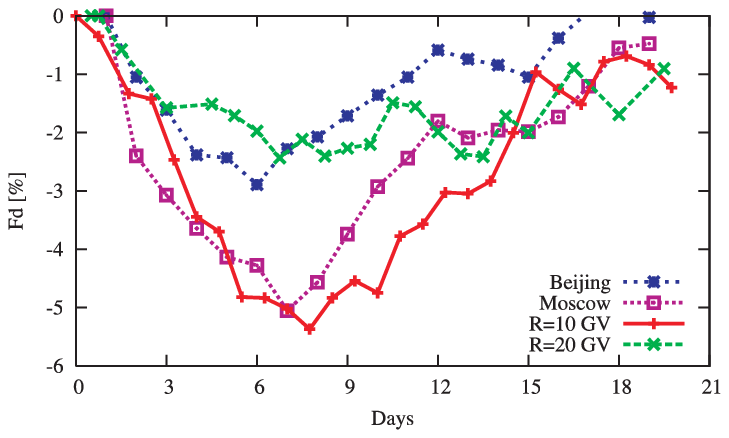}
\caption{\label{fig:ModelExper}Changes of the expected amplitudes of the Fd of the GCR intensity at the Earth orbit, for the rigidity of  10 and 20 GV based on the solutions of the backward SDEs in comparison with the GCR intensity registered by Moscow and Beijing neutron monitors during the Fd in March 2002.}
\end{minipage}
\end{figure}

In this paper we present the model of the recurrent Fd taking place due to established corotating heliolongitudinal disturbances in the interplanetary space. CIR passing the Earth gradually diminishes the diffusion at the Earth orbit, causing larger scattering of the GCR particles, and in effect fewer GCR particles reach the Earth. We simulate this process by the gradual decrease and then the increase of the diffusion coefficient at the Earth orbit with respect the heliolongitude. The diffusion coefficient $K_{II}$ of of cosmic ray particles has a form: $K_{II}=K_{0}\cdot K(r)\cdot K(R,\nu)$, where $K_{0}=10^{21}cm^{2}/s$, $K(r)=1+0.5\cdot (r/1 AU)$ and $K(R,\nu)=R^{2-\nu}$ . The exponent $\nu$ pronounce the increase of the HMF turbulence in the vicinity of space where the Fd is created (e.g. \cite{WA08, WA10}), and is taken as: $\nu = 0.8+ 0.2 sin(\varphi-90^{\circ})$ for $90^{\circ}\leq \varphi \leq 270^{\circ}$. We assume in the model the existence of the two dimensional spiral Parker's heliospheric magnetic field \cite{Parker58} implemented through the angle $\psi=arctan(-B_{\varphi}/B_{r})=arctan(\Omega \cdot r \cdot sin\theta/ U)$ in the 3D anisotropic diffusion tensor $K_{ij}$ of GCR particles \cite{Alania02}.\\
The expected changes of the GCR intensity for the rigidity of 10 and 20 GV during the simulated Fd in comparison with the profiles of the daily GCR intensities recorded by the two neutron monitors with different cut off rigidities in 18 March - 4 April 2002 presents Fig.~\ref{fig:ModelExper}. One can see that as is expected the amplitude of the Fd decreases for higher rigidities. One can see that the proposed model is in a good coincidence with the experimental data. Moreover, the model of the Fd obtained based on the solution of the SDE allows to reflect the stochastic character of the GCR particles  distribution in the heliosphere and present the pseudoparticle trajectory thorough the 3D heliosphere (Fig.~\ref{fig:Model}), which is not possible based on the solution of the Parker transport equation by the e.g. finite difference method (e.g.  \cite{WA08}).

\section*{Acknowledgments}
This work is supported by The Polish National Science Centre grant awarded by decision number DEC-2012/07/D/ST6/02488.

\end{document}